\documentclass[runningheads]{llncs}
\usepackage{graphicx}

\usepackage{amsmath}
\usepackage{amssymb}

\usepackage{bbding}
\usepackage{pifont}
\usepackage{wasysym}
\usepackage{utfsym}
\usepackage{fontawesome}
\usepackage{url}
\usepackage{subcaption}
\usepackage{mathtools}
\usepackage{color}
\usepackage{lipsum}

\usepackage{latexsym}
\usepackage{microtype}
\usepackage{mathrsfs}
\usepackage{graphicx}
\usepackage{algorithm}
\usepackage{algpseudocode}
\usepackage{multirow}
\usepackage{booktabs}
\usepackage{url}
\usepackage{adjustbox}
\usepackage{enumitem}
\usepackage{lipsum}
\usepackage{xcolor,colortbl}
\usepackage{array}
\usepackage{pifont}

\newcommand{\cmark}{\ding{51}}%
\newcommand{\xmark}{\ding{55}}%

\usepackage{hyperref}
\hypersetup{hidelinks,
	colorlinks=true,
	allcolors=black,
	pdfstartview=Fit,
	breaklinks=true}

\begin{document}
%

\title{Biomedical Image Reconstruction: A Survey}

%
\author{
Samuel Cahyawijaya \\
scahyawijaya@connect.ust.hk
}

%

%

\institute{HKUST}
%
%
\maketitle              

\begin{abstract}

Biomedical image reconstruction research has been developed for more than five decades, giving rise to various techniques such as central and filtered back projection. With the rise of deep learning technology, biomedical image reconstruction field has undergone a massive paradigm shift from analytical and iterative methods to deep learning methods To drive scientific discussion on advanced deep learning techniques for biomedical image reconstruction, a workshop focusing on deep biomedical image reconstruction, MLMIR, is introduced and is being held yearly since 2018. This survey paper is aimed  to provide basic knowledge in biomedical image reconstruction and the current research trend in biomedical image reconstruction based on the publications in MLMIR. This survey paper is intended for machine learning researchers to grasp a general understanding of the biomedical image reconstruction field and the current research trend in deep biomedical image reconstruction.

\end{abstract}

\section{Introduction}

Biomedical imaging provides scientists and physicians with an indispensable knowledge to understand, diagnose, and develop treatment of diseases inside the human body. Nonetheless, biomedical images acquired from imaging devices suffer from low signal-to-noise-ratio (SNR) and low contrast-to-noise ratio (CNR)~\cite{singh2012medical}. Various image reconstruction techniques have been developed to overcome these problems and to improve the quality
of images for better visual interpretation, understanding, and analysis. To overcome these limitations, various biomedical image reconstruction techniques have been developed which main goal is to obtain high-quality medical images with the lowest cost and risk to patients. These techniques have been widely adopted to enhance biomedical imaging acquired from various imaging devices, e.g., computed tomography (CT), magnetic resonance imaging (MRI), positron emission tomography (PET), radiography, mammography, ultrasonography, etc.

The aim of this survey is to help readers grasp the general understanding of the biomedical image reconstruction field along with its chronological progressions and recent trends. This survey covers the evolution of biomedical image reconstruction techniques starting from the classical one using various analytical and iterative approaches, dated back from a few decades ago, up to the recent paradigm shift of using deep learning techniques for biomedical image reconstruction. It is also worth noting that, unlike the classical techniques, the recent  deep learning biomedical image reconstruction techniques can be easily adopted across various application area.

The remainder of the survey is structured as follows. In Section \S\ref{sec:overview}, to help readers understand the basic concept of biomedical image reconstruction, the overview for both classical and deep learning techniques in biomedical image reconstruction techniques is described. 
To display the current progression and trend of biomedical image reconstruction, Section \S\ref{sec:evaluation} focuses on describing existing benchmarks, evaluation methods, and comparison of various classical and deep learning techniques in biomedical image reconstruction. In Section \S\ref{sec:discussion}, further discussion on the existing techniques, its limitations, and the potential improvements is described to further engage readers with the potential research direction of biomedical image reconstruction. And, in Section \S\ref{sec:conclusion}, a short summary of this survey is presented.

\section{Overview of Biomedical Image Reconstruction}
\label{sec:overview}

\vspace{-25pt}
\begin{figure}[!t]
    \centering
    \resizebox{\linewidth}{!}{
        \includegraphics{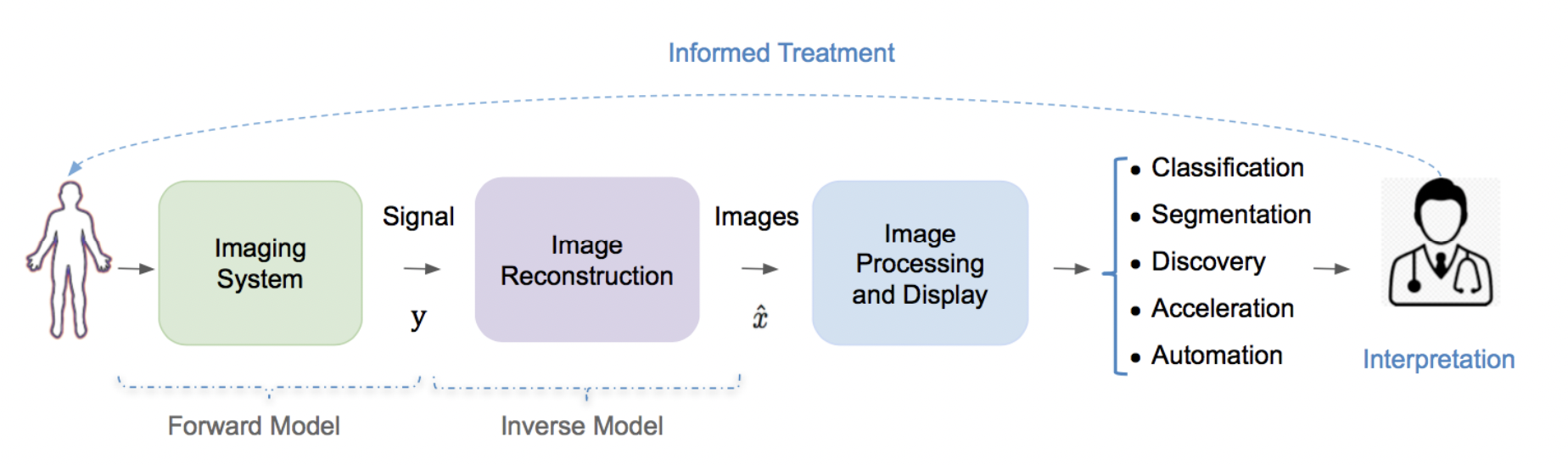}
    }
    \caption{The role of biomedical image reconstruction in biomedical imaging analysis. Image reconstruction is an inverse model which transforms the collected signals from the forward model into a meaningful image for further analysis.}
    \label{fig:overview}
\end{figure}

Biomedical image reconstruction is known as an inverse problem~\cite{benyedder2020deepreconstruction}. As shown in Figure~\ref{fig:overview}, a biomedical image reconstruction system takes a set of signals $y$ from a forward model $F(.)$ and the goal is to reconstruct the original structure of the object $\dot{x}$ in form of images. Mathematically, biomedical image reconstruction is called as an ill-posed problem as there is a low-dimensional measurements $y$ to determine the the high-dimensional target $\dot{x}$. Hence, there might be an infinite number of reconstruction images $\dot{x}$ that map to the same measurements $y$. Thus, one main challenge in any image reconstruction system is to find the best solution among a set of potential solutions~\cite{mccann2019foundationtodeep}, and one way to reduce the solution space is to use regularization through leveraging domain specific knowledge. Many solutions have been proposed for biomedical image reconstruction since more than four decades ago including various classical biomedical image reconstructions techniques and the more recent deep-learning-based biomedical image reconstruction techniques.  

\subsection{Classical Biomedical Image Reconstruction}

\begin{figure}[!t]
    \centering
    \resizebox{\linewidth}{!}{
    \includegraphics{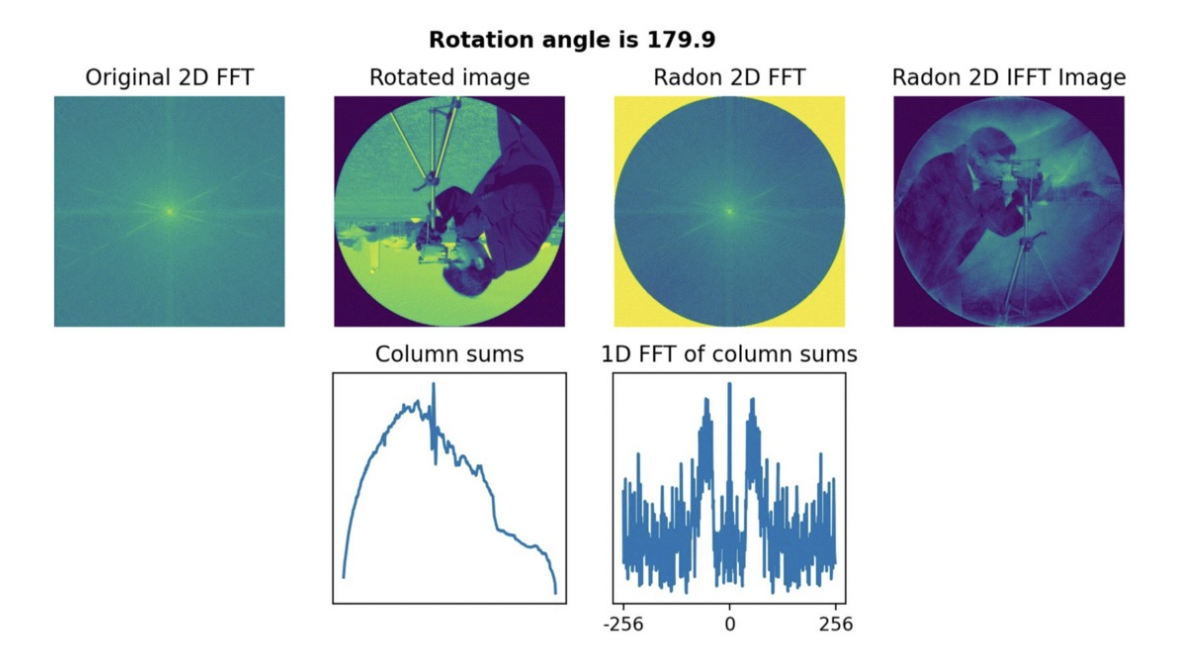}
    }
    \caption{An example of image reconstruction with Radon transform.}
    \label{fig:radon_transform}
\end{figure}

\begin{figure}[!t]
    \centering
    \resizebox{\linewidth}{!}{
    \includegraphics{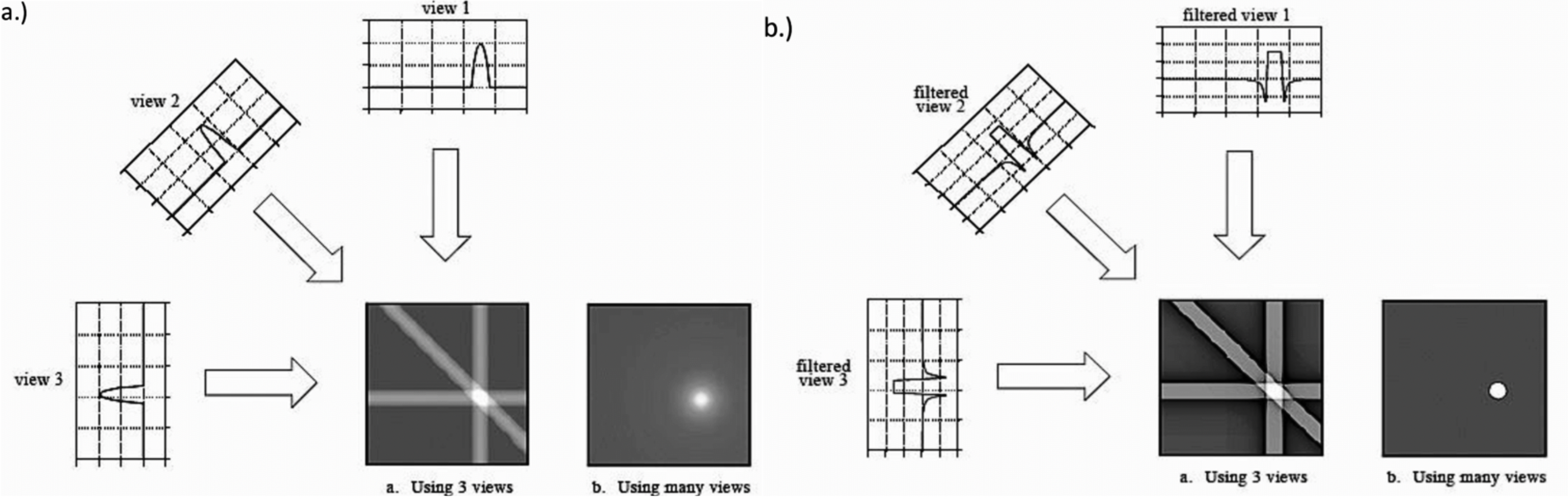}
    }
    \caption{An example of image reconstruction with a) back projection and b) filtered back projection.}
    \label{fig:filtered_back_projection}
\end{figure}

There are various classical biomedical image reconstruction techniques have been introduced which includes analytical methods (e.g., central slice theorem~\cite{kak1988principleofct} and filtered back projection~\cite{natterer1986mathofct,kak1988principleofct}) and iterative methods~\cite{nuyts1998iterative,griswold2002iterative,elbakri2002iterativestatistical,odille2010iterative,yin2009iterative,knopp2010iterative,dey2006iterativevariational,ivana2011dictionarylearning,yu2012oterative,nuyts2013iterative}. Central slice theorem specifies that the 1D Fourier transform of a projection taken at an angle $\theta$ is the same as the radial slice taken through the 2D Fourier domain of the object at the same angle~\cite{blackledge2005digitalimage}. Central slice theorem is applied in Radon transform which reconstructs object by repeating the process for multiple angle $\theta$ from 0 and $\pi$, and apply inverse Fourier transform to recover the object. Figure~\ref{fig:radon_transform} shows how central slice theorem is applied in Radon transform to reconstruct the image\footnote{For more intuitive explanation on Radon transform, readers are referred to \url{https://towardsdatascience.com/the-radon-transform-basic-principle-3179b33f773a}}. Back projection is the oldest and simplest projection reconstruction technique which reconstructs an object by taking multiple projections across different angles and sum them up~\cite{orrison1995bp}. Filtered back projection is an enhanced version of the back projection technique of which apply a convolution filter on each projection before summing them up to produce a high-contrast image~\cite{groenewald2017fbp}. Figure~\ref{fig:filtered_back_projection} shows the example of back projection and filtered back projection techniques. Although these analytical techniques are efficient, these techniques requires a proper and precise sampling from the imaging device which is often problematic.

To further improve the analytical techniques, iterative method that model the statistical and physical properties of the imaging device are introduced, such as algebraic reconstruction technique (ART)~\cite{gordon1974art}, simultaneous reconstruction technique (SIRT)~\cite{gilbert1972sirt}, simultaneous algebraic reconstruction technique (SART)~\cite{andersen1984sart}, maximum likelihood expectation maximization (ML-EM)~\cite{shepp1982mlem},  ordered subset expectation maximization (OSEM)~\cite{hudson1994osem}, maximum a posteriori~\cite{herman1992map}, total variation~\cite{dey2006iterativevariational,li2011iterativevariational}, and dictionary learning~\cite{ivana2011dictionarylearning,caballero2014dictionarylearning,ivana2011dictionarylearning,hu2016iterativedictionarylearning}. These techniques apply iterative optimization using different optimization criteria and generally produce better results compared to the analytical techniques. Despite having capability of modeling the properties of the imaging device, discrepancies between the model and physical factors are often occurred (e.g. inhomogeneous
magnetic fields, etc). These discrepancies between model and physical factors are often problematic and it becomes the main drawback of the iterative approach. 

\subsection{Deep Learning Biomedical Image Reconstruction}
\label{sec:deep}

In the recent years, biomedical image reconstruction undergoes a paradigm shift that is driven by advances in the deep learning technology~\cite{ahishakiye2021deepreconstruction,yaqub2022deepreconstructionmodal}. Whereas traditionally transform-based or optimization-based methods have dominated methods for image reconstruction, machine learning has opened up the opportunity for new data-driven approaches which have demonstrated a number of advantages over traditional approaches. In particular, deep learning techniques have shown significant potential for image reconstruction and offer interesting new approaches. Finally, deep learning approaches also offer to the possibility for application-specific image reconstruction, e.g. in motion-compensated cardiac or fetal imaging. Various deep learning architectures derived from generative adversarial network (GAN)~\cite{goodfellow2014gan} and U-Net~\cite{ronneberger2015unet} employing convolution neural network  (CNN)~\cite{lecun1989cnn}, recurrent neural network (RNN)~\cite{jordan1997rnn}, and transformer~\cite{vaswani2017transformer} have been proposed and achieve state-of-the-art performance in various biomedical image reconstruction tasks. Due to the uprising research works in the deep-learning-based biomedical image reconstruction, a workshop specified for biomedical image reconstruction, Machine Learning for Medical Image Reconstruction (MLMIR)~\cite{knoll2018mlmir,knoll2019mlmir,deeba2020mlmir,haq2021mlmir,haq2022mlmir}, is introduced and is being held yearly since 2018 to this year, publishing 84 novel deep learning techniques in biomedical image reconstruction.

\begin{table}[!t]
    \centering
    \begin{tabular}{
        >{\centering \arraybackslash}p{0.11\textwidth}|
        >{\centering \arraybackslash}p{0.1\textwidth}|
        >{\centering \arraybackslash}p{0.1\textwidth}|
        >{\centering \arraybackslash}p{0.1\textwidth}|
        >{\centering \arraybackslash}p{0.11\textwidth}|
        >{\centering \arraybackslash}p{0.11\textwidth}|
        >{\centering \arraybackslash}p{0.11\textwidth}|
        >{\centering \arraybackslash}p{0.1\textwidth}|
        >{\centering \arraybackslash}p{0.1\textwidth}
        }
        \toprule
            \textbf{MLMIR}  & 
            \multirow{2}{*}{\textbf{MRI}} & 
            \multirow{2}{*}{\textbf{CT}} & 
            \multirow{2}{*}{\textbf{PET}} & 
            \textbf{\small{Fluoro}} & 
            \textbf{Ultra} & 
            \textbf{Micro} & 
            \multirow{2}{*}{\textbf{MPI}} & 
            \multirow{2}{*}{\textbf{Others}} \\
            
            \textbf{Year} & & & & 
            \textbf{\footnotesize{scopy}} & \textbf{sound} & \textbf{scopy} & \\
        \midrule
            2018 & 9 & 3 & 0 & 0 & 1 & 0 & 0 & 4 \\
            2019 & 11$^{\dagger\dagger\dagger}$ & 6$^{\dagger\dagger}$ & 3$^{\dagger}$ & 0 & 2 & 3 & 0 & 2 \\
            2020 & 10 & 1$^{\star}$ & 1$^{\star}$ & 1 & 1 & 2 & 1 & 0 \\
            2021 & 10$^{\dagger}$ & 0 & 0 & 2 & 0 & 2$^{\dagger}$ & 0 & 0 \\
            2022 & 7 & 3 & 2 & 1 & 1 & 0 & 1 & 0 \\
        \bottomrule
    \end{tabular}
    \caption{Number of publications per modality in MLMIR workshop from 2018 to 2022. \textbf{Others} include mammography, sound-speed imaging, optoacoustic tomography, optical tomography, and general reconstruction approach. $^{\dagger}$ a work is evaluated on more than one modality. $^{\star}$ a work utilizes multimodal information.}
    \label{tab:modality_statistics}
\end{table}

\begin{table}[!t]
    \centering
    \begin{tabular}{
        >{\centering \arraybackslash}p{0.1\textwidth}|
        >{\centering \arraybackslash}p{0.1\textwidth}|
        >{\centering \arraybackslash}p{0.2\textwidth}|
        >{\centering \arraybackslash}p{0.1\textwidth}|
        >{\centering \arraybackslash}p{0.1\textwidth}|
        >{\centering \arraybackslash}p{0.1\textwidth}|
        >{\centering \arraybackslash}p{0.1\textwidth}|
        >{\centering \arraybackslash}p{0.1\textwidth}
        }
        \toprule
            \multicolumn{2}{c|}{\textbf{Derived From}} &
            \multirow{2}{*}{\textbf{Backbone}} & 
            \multirow{2}{*}{\textbf{2018}} & 
            \multirow{2}{*}{\textbf{2019}} & 
            \multirow{2}{*}{\textbf{2020}} & 
            \multirow{2}{*}{\textbf{2021}} & 
            \multirow{2}{*}{\textbf{2022}} \\
            \cmidrule{1-2}
            \textbf{U-Net} & \textbf{GAN} & & &  &  & \\
        \midrule
            \xmark & \xmark & CNN & 5 & 6 & 2 & 2 & 3 \\
            \xmark & \xmark & CNN \& RNN & 1 & 2 & 1 & 2 & 0 \\
            \xmark & \xmark & Transformer & - & - & - & 2 & 2 \\
            \cmark & \xmark & CNN & 6 & 4 & 6 & 4 & 4 \\
            \xmark & \cmark & CNN & 2 & 5 & 2 & 1 & 2 \\
            \cmark & \cmark & CNN & - & 2 & 1 & - & 1 \\
            \xmark & \xmark & VN & 1 & - & - & - & - \\
            \cmark & \xmark & VN & - & 1 & - & - & - \\
            \xmark & \xmark & Others & 1 & 4 & 3 & 2 & 3 \\
        \bottomrule
    \end{tabular}
    \caption{Number of publications per method novelty in MLMIR workshop from 2018 to 2022. \textbf{Others} include other model types (auto-encoder, dictionary learning, ML-EM, etc), other novelty (simulation tool, self-supervised learning methods, etc). VN denotes variational network.}
    \label{tab:method_statistics}
\end{table}

In order to grasp the general trend of deep biomedical image reconstruction, several key statistics from the publications in MLMIR are gathered\footnote{We realize that the statistics might not accurately capture the global trend in deep biomedical image reconstruction. Nevertheless, the statistics are enough to show the trend discussed of deep biomedical image reconstruction.}. Table~\ref{tab:modality_statistics} and Table~\ref{tab:method_statistics} shows the number of publication in MLMIR from 2018 to 2022 per modality and per architecture types, respectively. As shown in Table~\ref{tab:modality_statistics}, recent works in deep biomedical image reconstruction mostly focus on magnetic resonance imaging (MRI) and tomography imaging. A number of techniques also employ multimodal approach to improve the image reconstruction quality, while some others techniques can be applied to multiple modalities. Based to Table~\ref{tab:method_statistics}, we can see that in the recent years, most works in biomedical image reconstruction employ CNN backbone with hourglass architecture following U-Net~\cite{Dedmari2018,Hauptmann2018,Lin2019,Kasten2020,zhang2019apir,Zhang2020,Liu2020,Kim2022,Acar2022,Katare2022,Lee2022}, while the one without hourglass architecture has been declining~\cite{chaudhari2018,daltoso2019,zhang2019dict,aggarwal2019modl,choudhury2021,wang2020}. Other approaches that are commonly employed are CNN-based GAN~\cite{Murugesan2019,Cai2019,Ouyang2019,Yang2021,Kyung2022}. Since 2021, transformer-based architectures~\cite{Korkmaz2021,ma2021,yang2020learning,wang2022,Korkmaz2022} and various self-supervised pre-training approaches~\cite{Acar2021,MartnGonzlez2021,Huang2022} have been introduced for biomedical image reconstruction. Looking at the broader computer vision research field, these approaches are likely to gain more traction in the coming years. While other approaches including spatiotemporal modeling using CNN \& RNN modules~\cite{Oh2018,zhang2019rnn,Gadjimuradov2020,Pan2021} and iterative methods (e.g. variational network, dictionary learning, ML-EM, etc)~\cite{Vishnevskiy2018,Johnson2019,ktem2019,Green2019,zhang2019iterative} seem to lose their traction over years.

\section{Evaluating Biomedical Image Reconstruction}
\label{sec:evaluation}

\subsection{Datasets}

\begin{figure}[!t]
    \centering
    \resizebox{\linewidth}{!}{
       \includegraphics{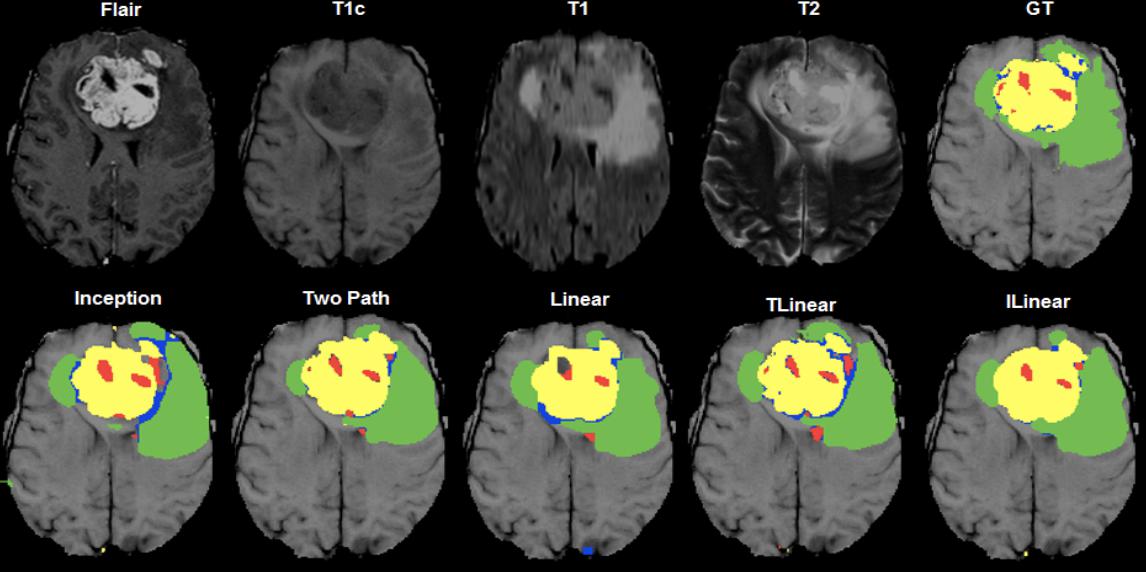}
    }
    \caption{Example of brain imaging data from BraTS challenge.}
    \label{fig:brats}
\end{figure}

There are various open datasets available for biomedical image reconstruction from various modalities.  Alzheimer’s Disease Neuroimaging Initiative (ADNI)~\cite{Jack2008adni} is a magnetic resonance (MR) neuroimaging dataset collected from 800 people aged 55-90 years. Approximately there are 200 MR data from cognitively normal elderly individuals, 400 MR data from individuals with MCI, and 200 MR data from individuals with early AD. Brain Tumour Segmentation (BraTS) challenge~\cite{Menze2015brats} focuses on evaluating state-of-the-art techniques for brain tumors segmentation in multimodal MRI. BraTS challenge has been held multiple times from 2013, 2015, 2017, 2018, 2020, 2021, and 2022 BraTS challenge. The Cancer Imaging Archive (TCIA)~\cite{Clark2013tcia} is a large-scale CT and PET/CT dataset from lung cancer patients. TCIA dataset also offers supporting evidence such as medical results, care specifics, genomics, and expert analyses. FastMRI~\cite{Knoll2020fastmri} is a MRI dataset consisting of 1,500 fully sampled knee MRI data and DICOM images from 10,000 clinical knees MRIs. OpenNeuro~\cite{Markiewicz2021openneuro} is a directory of open neuroimaging data obtained with various imaging modalities and protocols. The data in OpenNeuro is shared under a Creative Commons CC0 licence, which provides access to a large variety of brain imaging data. Autism Brain Imaging Data Exchange (ABIDE)~\cite{Martino2014abide} is an brain imaging dataset of infants with Autism Spectrum Disorder (ASD) and their controls, gathered from 24 different foreign brain imaging institutions. An example of the brain imaging dataset retrieved from the BraTS challenge is shown in Figure~\ref{fig:brats}.

\subsection{Evaluation Metrics}

\begin{figure}
    \centering
    \begin{minipage}{.5\linewidth}
        \centering
        \begingroup
        \includegraphics[width=\linewidth]{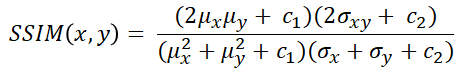}
        \endgroup
    \end{minipage}%
    \\
    \begin{minipage}{.3\linewidth}
        \centering
        \begingroup
        \includegraphics[width=\linewidth]{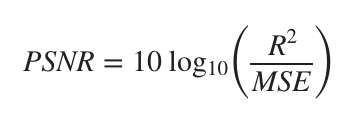}
        \endgroup
    \end{minipage}%
    \\
    \vspace{5pt}
    \begin{minipage}{0.3\linewidth}
        \centering
        \begingroup
        \includegraphics[width=\linewidth]{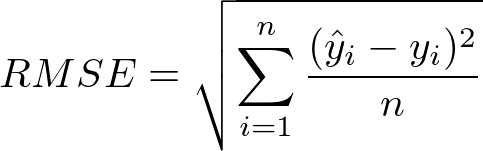}
        \endgroup
    \end{minipage}%
    \caption{Common evaluation metrics use in biomedical image reconstruction. \textbf{(top)} SSIM, \textbf{(middle)} PSNR, and \textbf{(bottom)} RMSE.}
    \label{fig:eval}
\end{figure}

For evaluating biomedical image reconstruction data, there are three commonly used evaluation metrics, i.e., structural similarity index (SSIM), peak signal-to-noise ratio (PSNR), root-mean-square error (RMSE). The definition of each evaluation metric is shown in Figure~\ref{fig:eval}

\section{Discussion}
\label{sec:discussion}

Biomedical image reconstruction field has been developed for more than five decades ago, starting from analytical techniques (i.e. central slice theorem and filtered back projection), iterative techniques (e.g., ML-EM, dictionary learning, ART, SIRT, etc), to the recent paradigm shift with deep learning approach. As explained in \S\ref{sec:deep}, various deep learning techniques have been explored over the last couple of years, with most of the work employing CNN-based model with hourglass architecture derived from U-Net. In recent years, there is a growing trend of self-supervised pre-training approaches and transformer-based model in the various fields such as natural language processing (NLP), audio signal processing, and computer vision (CV). Similar trend is also observed in biomedical image reconstruction field. In the coming years, we could expect that these three approaches (i.e., self-supervised pre-training, hourglass architecture, and transformer-based models) along with their combinations, might dominate the biomedical image reconstruction field. A number of works have already explored the potential of combining U-Net and Transformer, such as UCTransNet~\cite{wang2021uctransnet}, TransUNet~\cite{chen2021transunet}, UNETR~\cite{Hatamizadeh2021unetr}. One problem with transformer architecture is the quadratic space and time complexity on the attention mechanism, future work might address this problem to achieve a more efficient state-of-the-art biomedical image reconstruction method. Moreover, hourglass architecture also requires abundant of resources for computing high dimensional data, such as 3D volumetric data or 4D (3D volumetric + time) data. Combining hourglass architecture with transformer will likely to yield higher computational cost, and explorations to mitigate this problem is urgently needed. 

\section{Conclusion}
\label{sec:conclusion}

This survey paper explains the overview of the research in the biomedical image reconstruction field. The role of biomedical image reconstruction, datasets, evaluation metrics, and various techniques for biomedical image reconstruction have been elaborated, starting from the classical biomedical image reconstruction techniques using analytical methods and iterative methods, which is then followed by a paradigm shift into deep learning techniques. The trend of recent deep learning approaches for biomedical image reconstruction has also been elaborated with hourglass architecture, transformer architecture, and self-supervised pre-training as three currently trending approaches in biomedical image reconstruction. Future work might consider the novelty on these three approaches along with theirs combination, while at the same time considering the performance-efficiency trade-off of such approaches in performing real-case biomedical image reconstruction.


\bibliographystyle{splncs04}
\bibliography{reference}

\end{document}